\documentclass[12pt,preprint]{aastex}

\slugcomment{Not to appear in Nonlearned J., 45.}

\shorttitle{Photometric and spectroscopic light curves of paleo-Earths}
\shortauthors{Sanrom\'a & Pall\'e & Garc\'ia Mun\~oz}

\usepackage[normalem]{ulem}

\begin{document}




\title{On the effects of the evolution of microbial mats and land plants on the Earth as a planet. Photometric and spectroscopic light curves of paleo-Earths}







\author{E. Sanrom\'a\altaffilmark{1,2}, E. Pall\'e\altaffilmark{1,2} and A. Garc\'ia Mun\~oz\altaffilmark{1,2}}

\affil{Instituto de Astrof\'isica de Canarias (IAC), V\'ia L\'actea s/n 38200, La Laguna, Spain}

\email{mesr@iac.es}





\altaffiltext{2}{Departamento de Astrof\'isica, Universidad de La Laguna, Spain}






\begin{abstract}

Understanding the spectral and photometric variability of the Earth and the rest of the solar system planets has become 
of the utmost importance for the future characterization of rocky exoplanets. As this is not only interesting at present 
times but also along the planetary evolution, we studied the effect that the evolution of microbial mats and plants over 
land has had on the way our planet looks from afar. As life evolved, continental surfaces changed gradually and non-uniformly 
from deserts through microbial mats to land plants, modifying the reflective properties of the ground and most probably the 
distribution of moisture and cloudiness. Here, we used a radiative transfer model of the Earth, together with geological 
paleo-records of the continental distribution and a reconstructed cloud distribution, to simulate the visible and near-IR 
radiation reflected by our planet as a function of Earth's rotation. We found that the evolution from deserts to microbial 
mats and to land plants produces detectable changes in the globally averaged Earth's reflectance. The variability of each 
surface type is located in different bands and can induce reflectance changes of up to 40\% in period of hours. We conclude 
that using photometric observations of an Earth-like planet at different photometric bands, it would be possible to 
discriminate between different surface types. While recent literature propose the red edge feature of vegetation near 0.7 $\mu$$m$ 
as a signature for land plants, observations in near-IR bands can be equally or even better suited for this purpose.

\end{abstract}






\section{INTRODUCTION}

In the last decades, more than 850 exoplanets have been detected outside the solar system, while thousands of 
potential planet candidates from the Kepler mission are waiting for confirmation.
Even though most of the discovered exoplanets are gas giants, as the larger planets are easier to detect than 
the smaller rocky ones, evolving observational capabilities have already allowed us to discover tens of planets in the 
super-Earth mass range (e.g. \citealt{Udr07, Cha09, Pep11, Bor12}), 
some of them probably lying within the habitable zone of their stars (\citealt{Bor12}). Moreover, some 
Earth-sized, and even smaller, exoplanets have already been reported in the literature (\citealt{Fre12,Mui12}). 
Indeed, early statistics indicate that about 62\% of the Milky Way stars may host a super-Earth (\citealt{Cas12}). 
Thus, one can confidently expect that true Earth analogues will be discovered in large numbers in the near future.

To be prepared for the characterization of future exoearth detections, the exploration of our own solar system and its planets is essential. 
This will allow us to test our theories and models, enabling more accurate determinations, and characterization 
of the exoplanets' atmospheres and surfaces. In particular, observation of the solar system rocky planets, 
including Earth, will be key for the search for life elsewhere.  

Over the last years, a variety of studies, both observational and theoretical, to determine how the Earth 
would look like to an extrasolar observer have been carried out. One of the observational approaches has 
been to observe the Earthshine, i.e., the sunlight reflected by Earth via the dark side of the moon. The 
visible spectrum of the Earthshine has been studied by several authors (\citealt{Goo01,Woo02,Qiu03,Pal03,Pal04}), 
while more recent studies have extended these observations to the 
near-infrared (\citealt{Tur06}) and to the near-UV (\citealt{Ham06}).

Several authors have also attempted to measure the characteristics of the reflected spectrum and the 
enhancement of Earth's reflectance at 700 nm due to the presence of vegetation, known as red-edge, 
directly (\citealt{Arn02,Woo02,Sea05,Mon06,Ham06}), 
and also by using simulations (\citealt{Tin06a,Tin06b,Mon06}). 
The red-edge has been proposed as a possible biomarker in Earth-like planets (\citealt{Tin06c,Kia07a,Kia07b}). 
Furthermore, \citet{Pal08} determined that the light scattered by the Earth as a function of time contains 
sufficient information, even with the presence of clouds, to accurately measure Earth's rotation 
period. More recently, \citet{Ste12} have also studied the use of the linear polarization 
content of the Earthshine to detect biosignatures, and were able to determine the fraction of 
clouds, oceans, and even vegetation.

Another approach has been to analyze Earth's observations from remote-sensing platforms 
(e.g., \citealt{Cow09,Cow11,Rob11}). These kinds of observations have 
also allowed the possibility to reconstruct the continental distribution of our own planet 
from scattered light curves. \citet{Cow09} performed principal components analysis in order 
to reconstruct surface features from the EPOXI data. \citet{Kaw10,Kaw11} and \citet{Fuj12} proposed an inversion 
technique which enables to sketch a two-dimensional albedo map from annual variations of the disk-averaged 
scattered light. In addition, \citet{Oak09} attempted to reproduce a longitudinal map of the Earth 
from simulated photometric data by using the difference in reflectivity between land and oceans. 
Other authors have also studied what the color of an extreme Earth-like planet
might be by utilizing filter photometry (\citealt{Heg13}), while others have studied the disk-averaged spectra
of cryptic photosynthesis worlds (\citealt{Coc09}).

However, it is unlikely that, even if we were to find an Earth-twin, planet will be at an evolutionary 
stage similar to the Earth is today. On the contrary, extrasolar planets are expected 
to exhibit a wide range of ages and evolutionary stages. Because of this, it is of interest not only to 
use our own planet, as it is today, as an exemplar case, but also at different epochs (\citealt{Kal07}).

The development of advanced plants is believed to have taken place on Earth during the Late Ordovician, about 450 Ma 
ago, albeit fungi, algae, and lichens may have greened many land areas before then (\citealt{Gra85}). 
Microbial mats are 
multilayered sheets of microorganisms generally composed of both Prokaryotes and Eukaryotes, 
being able to reach a thickness of a few centimeters. The time when microbial mats 
appeared on the Earth surface is still not clear, but prior to the evolution of algae and land
plants on early Earth, photosynthetic microbial mats probably were among the major forms of life on our planet. 
Microbial mats are found in the fossil record as early as 3.5 billion years ago.  
Later, when advanced plants and animals evolved, extensive microbial mats became rarer, 
but they are still presented in our planet in many ecosystems (\citealt{Sec10}). Even today, 
they still persist in special environments such as thermal springs, high salinity environments, and sulfur springs.

\begin{figure*}
   \centering
   \includegraphics[width=9.3cm]{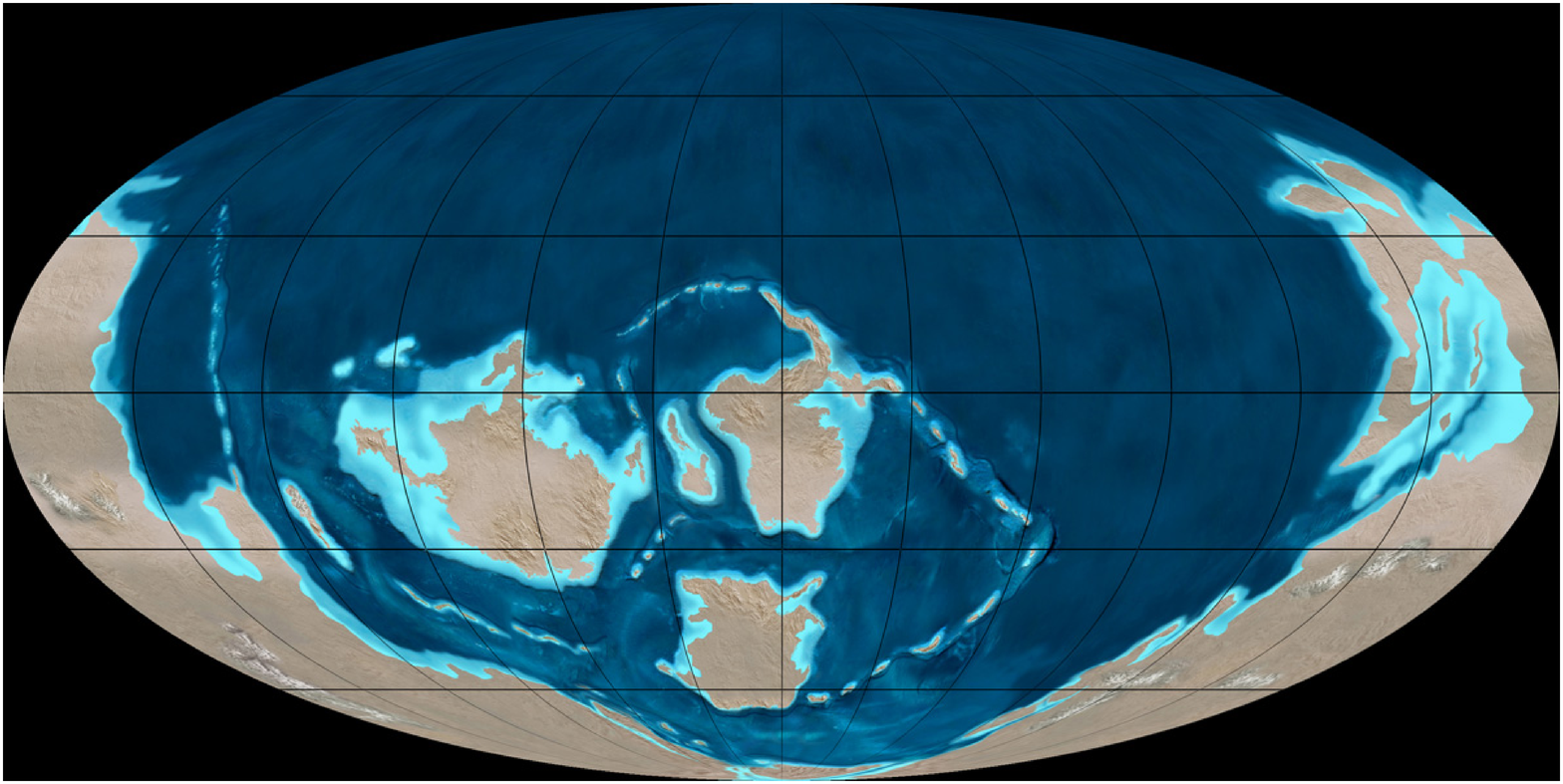}
   \caption{Geologic map of the Earth during the Late Cambrian (500 Ma ago). In our simulations the meridian line
	    crossing the image center has been taken as longitude 0. Image credit: Ron Blakey. }                                                                                              
              \label{mapa}%
    \end{figure*}

In this paper, we aimed to discern the effect that the evolution of life over land might have had on the way 
our planet would look like to a remote extraterrestrial observer. To this end, we have simulated both visible and 
near-IR disk-integrated spectra of our planet, considering as scenario of such simulations the Earth during 
the Late Cambrian (Figure \ref{mapa}), the period during which the development of the first plants is believed to have taken place. Thus, it is 
clear that as life evolved over land, Earth's continental masses must have changed gradually going 
through different states that go from deserts, to microbial mats, to evolved plants.

%__________________________________________________________________

\section{MODEL DESCRIPTION}

With the aim of deriving disk-averaged spectra of the Earth 500 Ma ago, at any viewing and illumination 
geometry, and for several surface types covering the continental crust, we have generated a database of 
one-dimensional synthetic radiance of the Earth, i.e., we have calculated synthetic spectra for a variety 
of surface and cloud types, and for several viewing and illumination angles. To do that, we have used a 
line-by-line  radiative transfer algorithm, based on the DISORT\footnote{ftp://climate1.gsfc.nasa.gov} 
(Discrete Ordinates Radiative Transfer Program for a Multi-Layered Plane-Parallel Medium) code 
(\citealt{Sta88}).

Our radiative transfer model (RTM) uses profiles of temperature and atmospheric composition, 
spectral albedos of each surface type, and cloudiness information as input data 
for the calculations (see subsequent subsections for a detailed description of inputs). 
The code considers Earth's atmosphere as plane-parallel, and its radiative properties 
are prescribed on a grid of contiguous spectral bins. This spectral grid is unevenly spaced, 
and designed to resolve the rapid variations in the radiative properties near molecular absorption 
lines. Each spectrum is calculated at very high spectral resolution, with no less than three points
 per Doppler width.  Positions, intensities and lineshape parameters for molecular absorption bands 
are taken from HITRAN2008 (\citealt{Rot09}). Only a single angle of incidence and 10 angles of 
reflection can be used for each model run. Finally, each spectrum is degraded to a lower resolution 
for storage purposes ($R=10,000$). We have run this radiative code using input parameters covering a broad range of viewing and 
illumination angles, surface and cloud types, atmospheric profiles, and aerosol concentrations, 
leading to the generation of a database containing about 7000 spectra. The RTM is essentially an 
extension of the RTM for transits described in \citet{Gar11} and \citet{Gar12} to a viewing geometry for which 
the light reaching the observer has been reflected at the planet. The RTM is also capable of modeling the 
planet's emission of thermal radiation, although that possibility is not explored here.

Once the spectral library was generated, we developed a computer code to calculate the disk-integrated 
irradiance of our planet for given sub-solar and sub-observer points, a surface map, and a cloudiness 
distribution. At any given date the code calculates both the fraction of the planet visible to the observer 
and the illuminated fraction of the Earth from the location of the observer. The planets is subdivided in a 
64x32 pixel (longitude by latitude) grid. This grid resolution offers the best balance between 
computational cost and accuracy. For each grid a radiance spectrum from the aforementioned 
database is assigned to each detectable and illuminated surface's pixel taking into account the surface type, 
and the percentage cloud type and amount of such pixel. Note that the full spectral database was generated 
only for 10 solar angles and for 10 observer angles (85$^\circ$, 80$^\circ$, 75$^\circ$, 70$^\circ$, 65$^\circ$, 
55$^\circ$, 45$^\circ$, 30$^\circ$, 15$^\circ$, and 0$^\circ$ for both angles). Thus, radiance values for arbitrary 
solar and observer angles were interpolated into the angles from the calculated radiances.

Finally, to get the disk-averaged spectrum  of the ancient Earth, we have weighted each pixel's 
spectrum by their solid angle and we have integrated over the whole Earth.

\subsection{Atmospheric Properties}

Information about temperature and distribution of atmospheric gases were taken from FSCATM 
(\citealt{Gal83}). We have considered five atmospheric profiles models: tropical, midlatitud summer, 
midlatitud winter, subartic summer, and subartic winter. These atmospheric profiles include 
mixing ratios of the most significant molecules in Earth's atmosphere: $H_{2}O$, $CO_{2}$, $O_{3}$, 
$CH_{4}$, $O_{2}$ and $N_{2}$. These properties are prescribed  into 33 uneven layers, which go from 0 to 
100 Km height, being the spacing between layers of 1 km near the bottom of the atmosphere, and 5 km or more above 
25 km height.

Note that Earth's atmosphere 500 Ma ago can be considered similar to the present one since, on average, the same 
atmospheric composition and mean averaged temperature have existed during this period (\citealt{Har78,Kas02}). 
This is a necessary condition for our simulations to hold as a valid approximation.

\subsection{Surface Distribution and Albedos}

\begin{figure*}
   \centering
   \includegraphics{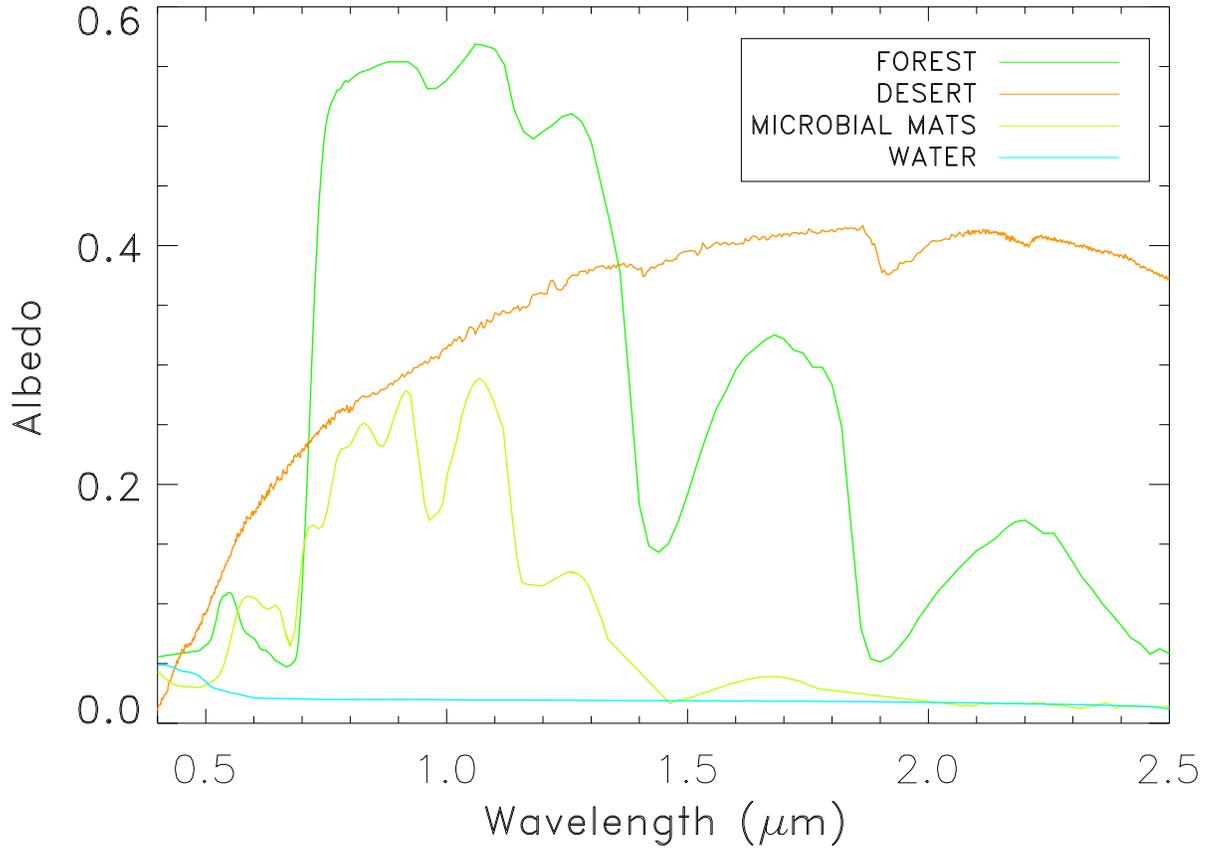}
   \caption{Spectral reflectance of the four surface types used in our model for the Earth 500Ma ago: forest (green), 
	    desert (orange), microbial mats (light green) and water (blue).}
              \label{albedos}%
    \end{figure*}

To carry out our goal of finding out how the appearance of life over land could affect Earth's reflectance properties,
we have considered four different continental land types: water, desert, microbial mats and forest. 
Figure \ref{albedos} shows the wavelength-dependent albedos of each surface type according to the 
ASTER Spectral Library\footnote{http://speclib.jpl.nasa.gov} and the USGS Digital 
Spectral Library\footnote{http://speclab.cr.usgs.gov/spectral-lib.html}. 

The continental distribution during the Late Cambrian has been taken from Ron Blakey's Web site\footnote{http://jan.ucc.nau.edu}. 
Surface maps of our planet in that epoch are available online. The Earth geologic information has been regridded 
into the 64x32 pixel grid used by our model.

\subsection{Cloud Distribution and Optical Properties}

In our model, the spatial distribution of clouds was taken from the International Satellite Cloud Climatology 
Project (ISCCP; \citealt{Ros96})  cloud climatology. With the aim of getting information about 
the global distribution of clouds over different surface types,  and in order to reconstruct 
the possible cloud distribution of the Late Cambrian (500 Ma ago), we have proceeded in the same 
way as in \citet{San12} but here, instead of using only the total cloud amount information, 
we have used information on three separate cloud layers: low (1000-680 Mb), mid (680-440 Mb) and high cloud (440-30 Mb) data. 

% The optical properties of each cloud types, scattering and absorption coefficients, were obtained from the 
% Optical Properties of Aerosols and Clouds (OPAC) data base (\citealt{Hes98}).

The optical properties of each cloud type, wavelength-dependent scattering and absorption coefficients, 
and the asymmetry parameter, were obtained from the Optical Properties of Aerosols and Clouds (OPAC) data base (\citealt{Hes98}).
We have considered physical cloud thicknesses of 1 km, and we have assumed that the scattering phase function
is described by the Henyey-Greenstein equation inside clouds, and by the Rayleigh scattering function outside them. 

We calculated the spatial distribution of these three cloud types according to the surface type 
lying beneath. To perform this classification, we used real geographical information about the different 
types of surfaces and vegetation in the present Earth, available from  the  ISCCP 
Web site\footnote{http://isccp.giss.nasa.gov}. Although the ISCCP classifies the surfaces as water, 
rain-forest, deciduous-forest, evergreen-forest, grassland, tundra, shrub-land, desert, and ice, 
we have only distinguished between four surface types: water, desert, ice, and vegetation, where the 
latest is  defined as the sum of rain-forest, deciduous-forest, evergreen-forest, grassland, tundra, 
or shrub-land. We then computed the mean cloudiness at each latitude point, obtaining the empirical 
relationships between the amount of clouds, surface type,  and latitude for low, mid, and high clouds 
(see \citet{San12} for details on the method). These relationships allow us to reconstruct 
the possible cloudiness distribution of the Earth 500 Ma ago.

It is worth noting that we do not have any empirical information about how clouds behave over extended 
microbial mat surfaces. Thus in our calculation we have used both the cloudiness information corresponding 
to deserts and vegetation. Probably the answer is somewhere in between, as microbial mats will increase 
transpiration and moisture compared to bare land. However, we will show how the choice of cloud amounts does 
not substantially affect the results.

% \subsection{Solar spectrum}
% 
% For the upper boundary condition of the model, a spectral distribution of the solar flux at the top  of the 
% atmosphere is needed. In this work, the solar spectrum was taken from the 1985 Wehrli Standard  Extraterrestrial 
% Solar Irradiance Spectrum\footnote{http://rredc.nrel.gov/solar/spectra/am0}.

%__________________________________________________________________

\section{SPECTRA AND LIGHT CURVES OF EARTH 500MA AGO}

\subsection{Spectral Models}
The development of advanced land plants is believed to have taken place during the Late Ordovician, about 450 Ma ago, 
albeit fungi, algae, and lichens may have greened many land areas long before (\citealt{Gra85}). Thus, 
to determine the impact that such changes could cause in the appearance of the ancient Earth's spectrum,  
we have run our model for four separate scenarios. First of all, to illustrate the spectrum properties of a 
planet without life on its surface, we have considered that  the continental crust of the Earth 500 Ma ago 
was entirely desert. While this might not be entirely true, it certainly was true at some point earlier in Earth's history, 
with an unknown continental distribution. 
Here, we chose to keep the Late Cambrian continental structure in order to isolate the spectral changes due to land 
scenery changes from those introduced by changing continental distribution (\citealt{San12}). Then, we have regarded a case where continents were covered by microbial mats. 
Here, two separate simulations were run, with the cloud types distribution over land corresponding to desert and to vegetated areas. 
We have also considered the case where the continental crust during the Late Cambrian was entirely covered by 
evolved plants, like those that dominate the continental surface today. Finally, in Section \ref{sec_photometric} we 
have also considered more realistic cases with 50\% mixtures between different surface types. 
In all cases ocean areas remained the same, and a three-layer cloud distribution over the whole planet is applied. 

\begin{figure*}
   \centering
   \includegraphics{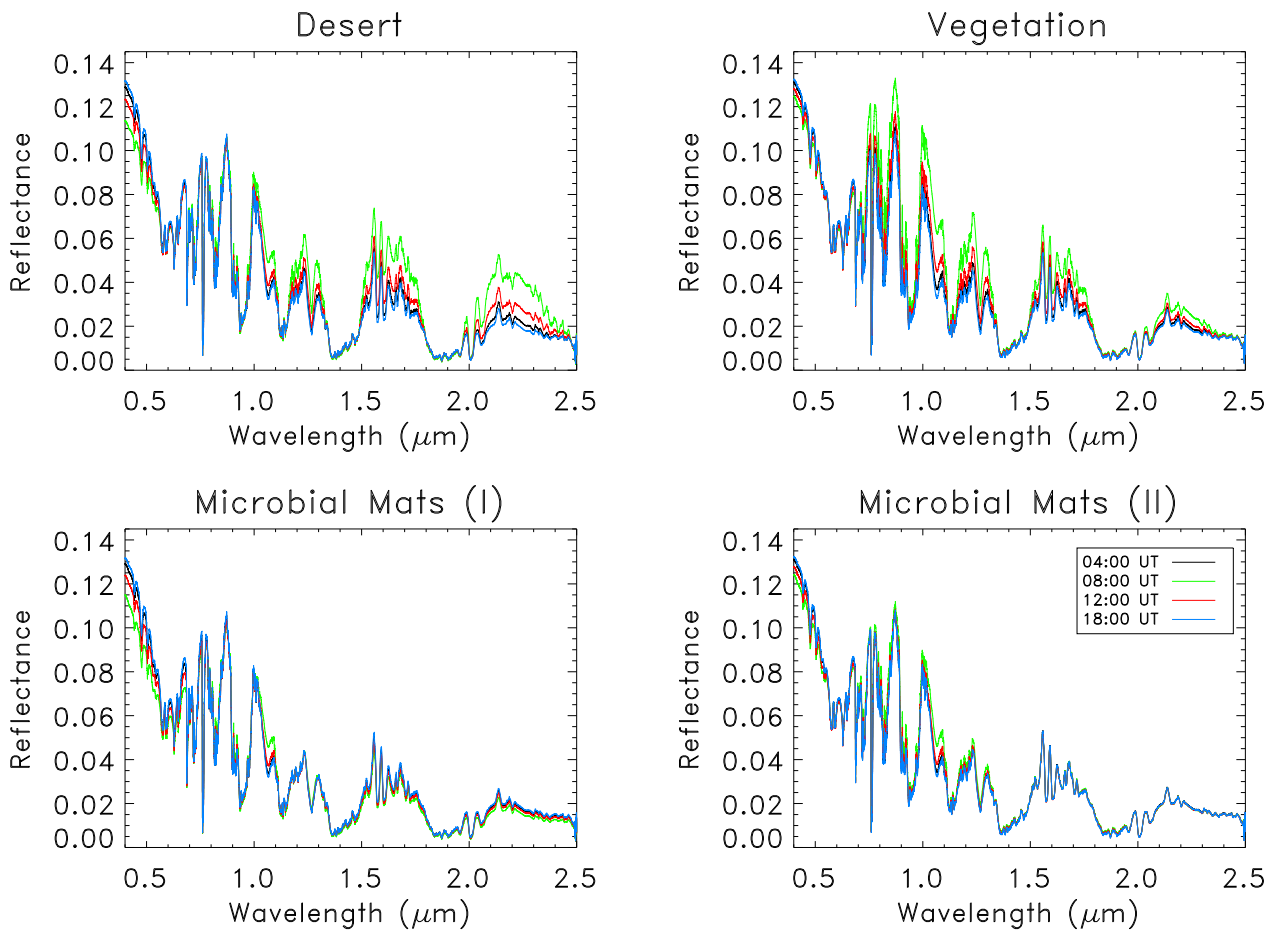}
   \caption{Earth's reflectance 500 Ma ago, taken as $\pi$ times the disk-averaged radiance divided by 
	    the solar flux, in the VIS-NIR, 
	    for our four cases in which the continents are covered by: deserts, vegetation, microbial mats with the 
	    cloudiness information corresponding to desert (I), 
	    and microbial mats with the cloudiness information corresponding to vegetation (II). The Earth is 
	    viewed at a phase angle of 90$^\circ$ and 
	    the different spectra follow their diurnal rotation. Both, the observer and the Sun are 
	    located over the equator. The spectra have been smoothed with a 100 point running mean for illustration purposes.} 
              \label{500maespectro}%
    \end{figure*}

Figure \ref{500maespectro} shows synthetic disk-integrated spectra of the Earth 500 Ma ago covering the spectral 
ranges between 0.4 and 2.5 $\mu$$m$, at different times of the day, for each of our four cases where continents are 
totally covered by i) deserts, ii) vegetation, iii) microbial mats with the cloudiness information 
corresponding to deserts, and iv) microbial mats with the clouds corresponding to vegetation. 
In all cases throughout this paper, the observer and the Sun are both located over the planet's equator in 
such a way that the observer is looking at a quarter-illuminated planet (phase angle $90^\circ$). For an extrasolar planet, 
the maximum angular separation from the parent star along the orbit occurs at phase $90^\circ$, as defined from the 
observer's position. Thus, this is the more relevant geometry for future exoplanet studies.

One can note that as continents come in and out of the field of view, the 
global Earth's spectra change. At 8:00 UT, when the continental presence is at maximum, the changes are more 
dramatic, in contrast to the spectra at 18:00 UT where differences between the four scenarios cannot be 
appreciated. That is related with the percentage of land that were illuminated and visible at each time. 
These percentages are approximately 24\%, 46\%, 36\%, and 6\% for 4:00, 8:00, 12:00, and 18:00 UT, respectively.

In order to show such intensity changes in terms of the strength of the vegetation's red-edge, we have calculated 
the ratio between the 0.740-0.750 $\mu$$m$ and the 0.678-0.682 $\mu$$m$ spectral ranges (Table \ref{tbl-1}). 
The choice of these spectral regions was made according to \citet{Mon06}. As expected, at 18:00 UT (6\% of land) this 
rate is almost the same in our four cases, while at 08:00 UT (46\% of land) the maximum is reached in the vegetation 
case where the red-edge is more pronounced.

For a planet covered by bare dirt, Figure \ref{500maespectro} illustrates some diurnal temporal variability, 
which is small in the visible range, but increases toward redder wavelengths. \citet{Gom12} studied 
the mid-IR emission of the planet in order to derive the rotational period. They found that the 
signature of the emission from the Sahara desert, among other, can be seen from space and determines a clear 
maximum in emission. Thus, it is probable that this variability would be even larger in the mid-IR.

In the case of a planet with continents covered by microbial mats, the spectra do not seem to vary much along 
the planet rotation, despite the fact of having a large and well differentiated land mass. The impact of using two different clouds  layers in the model (desert or vegetation) is almost negligible, being the former just a little bit more variable. 

On the contrary the spectrum of a planet covered entirely by plants varies significantly along the day, 
with the maximum variance occurring in the 
visible and near-IR. In particular the variability along the red-edge feature (at 700 $nm$) is 
large, and it extends all the way into the near-IR.

\subsection{Photometric Light-Curves}
\label{sec_photometric}

Observing a detailed reflectance spectrum of an exoplanet would no doubt be a major advantage toward characterizing its atmospheric 
composition and surface features. However, given the low 
signal to noise ratio scenarios that are expected for such observations, spectral information might be difficult 
to retrieve, especially at high enough temporal resolution to allow sampling of the planet's rotation. 
Broadband photometric 
observations are perhaps a more realistic scenario. In order to simulate such photometric observations, 
we have convolved our modeled spectra with standard visible and near-infrared photometric filters, namely, 
B,V, R, I, z, J, H, and K. 

\begin{figure*}
   \centering
   \includegraphics{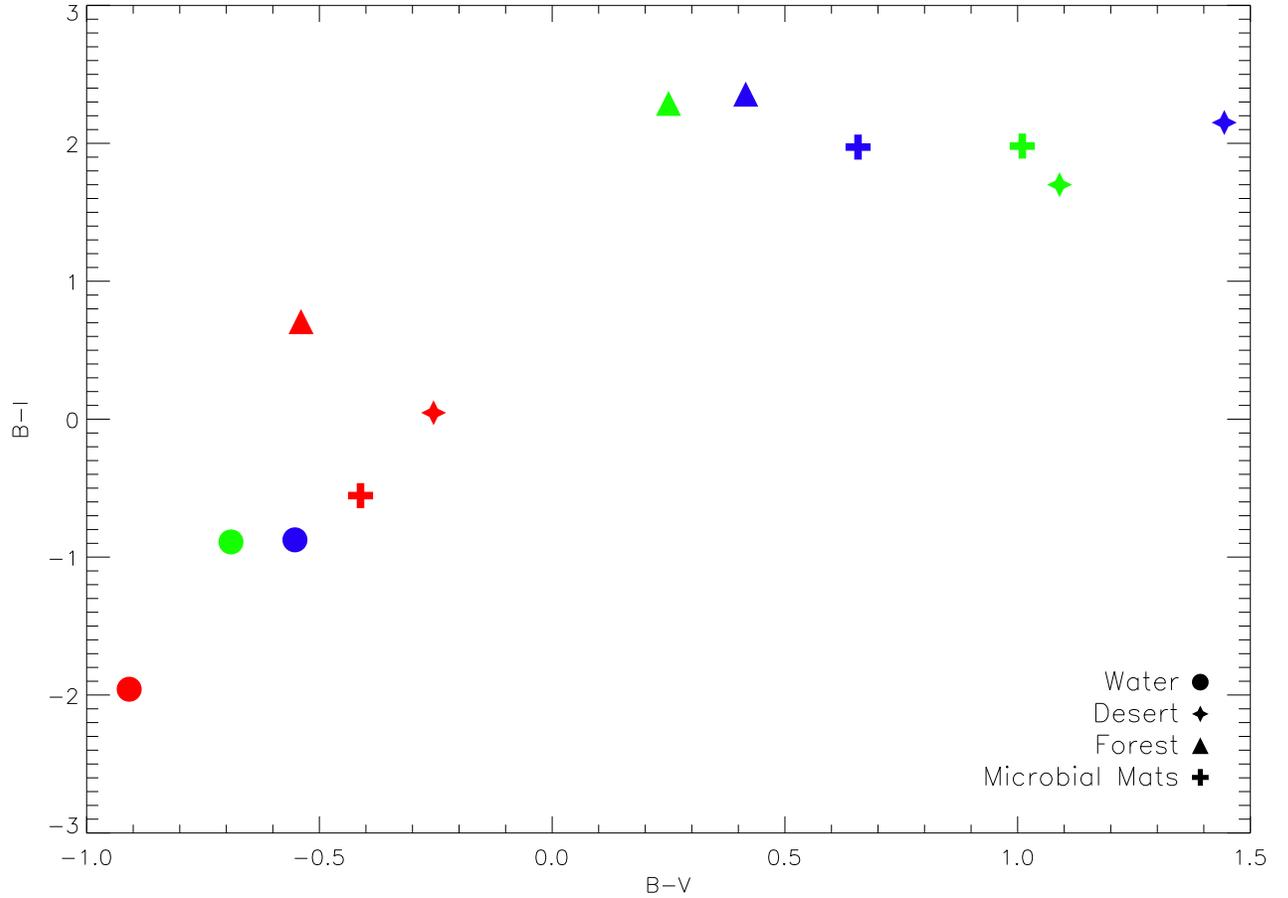}

   \caption{Color-color diagram of homogeneous test planets. Symbols denote planets whose surface is totally covered by water
	   (circles), deserts (stars), vegetation (triangles), and microbial mats (crosses). The values are shown
	   for atmosphere-less planets (this work in blue and \citealt{Heg13} in green), and for planets 
	   with an earth-like atmosphere (red). Notice that adding an atmosphere to the model reduces considerably the variability of 
	   the colors of the different planets.}
              \label{HKcomparison}%
    \end{figure*}

Figure \ref{HKcomparison} shows 
the B-V versus B-I color-color diagram of test planets fully covered by one specific surface type, i.e., planets totally covered 
by water, desert, vegetation, and microbial mats, with and without atmosphere (blue and red symbols, respectively). As can 
be seen, the addition of an atmosphere in our models moves the position of the planets in the color-color diagram significantly, 
reducing the color spread of the different surface types along the diagram. 

\begin{figure*}
   \begin{center}
   \includegraphics[width=5.3in]{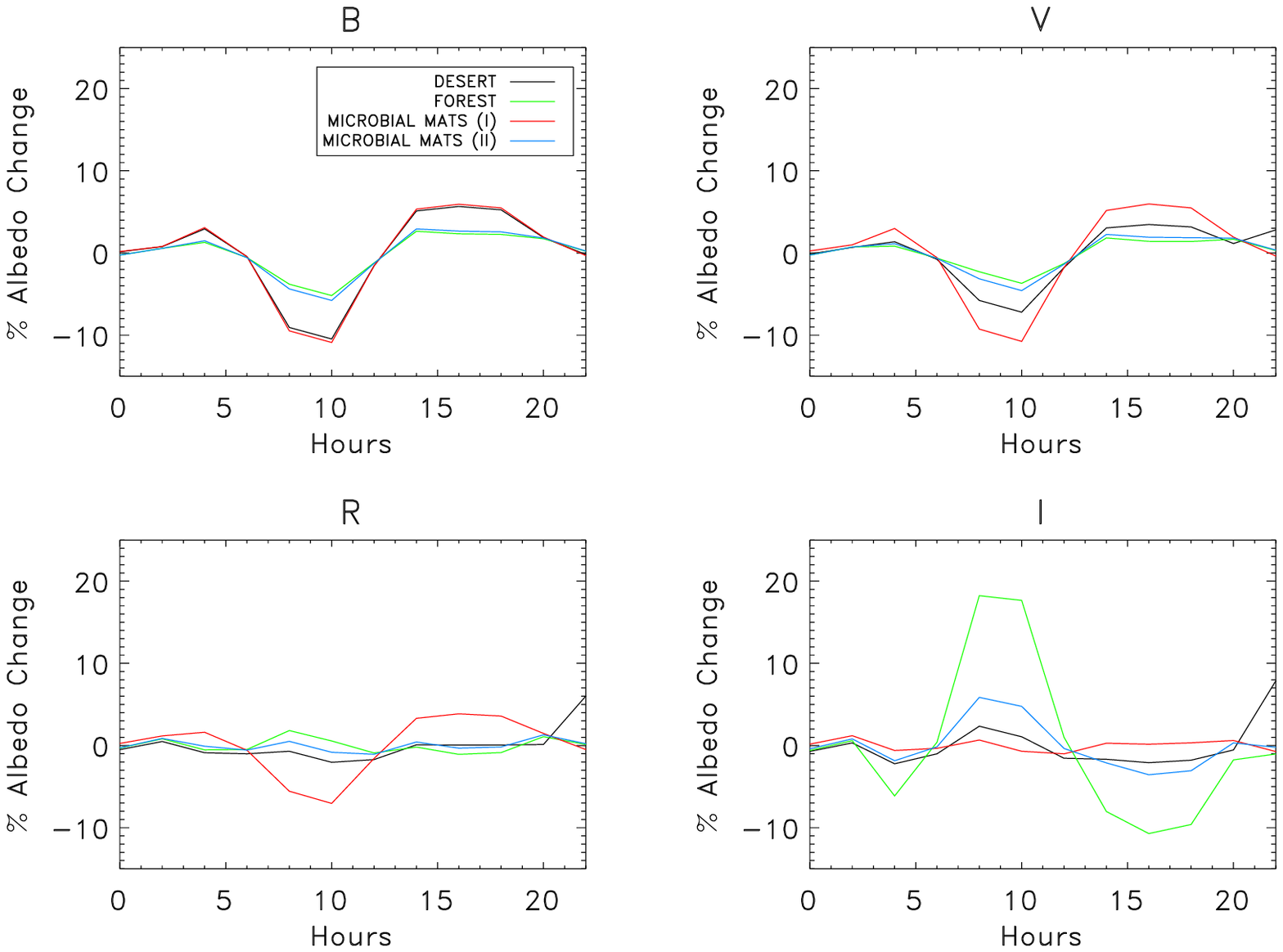}\\
   \includegraphics[width=5.3in]{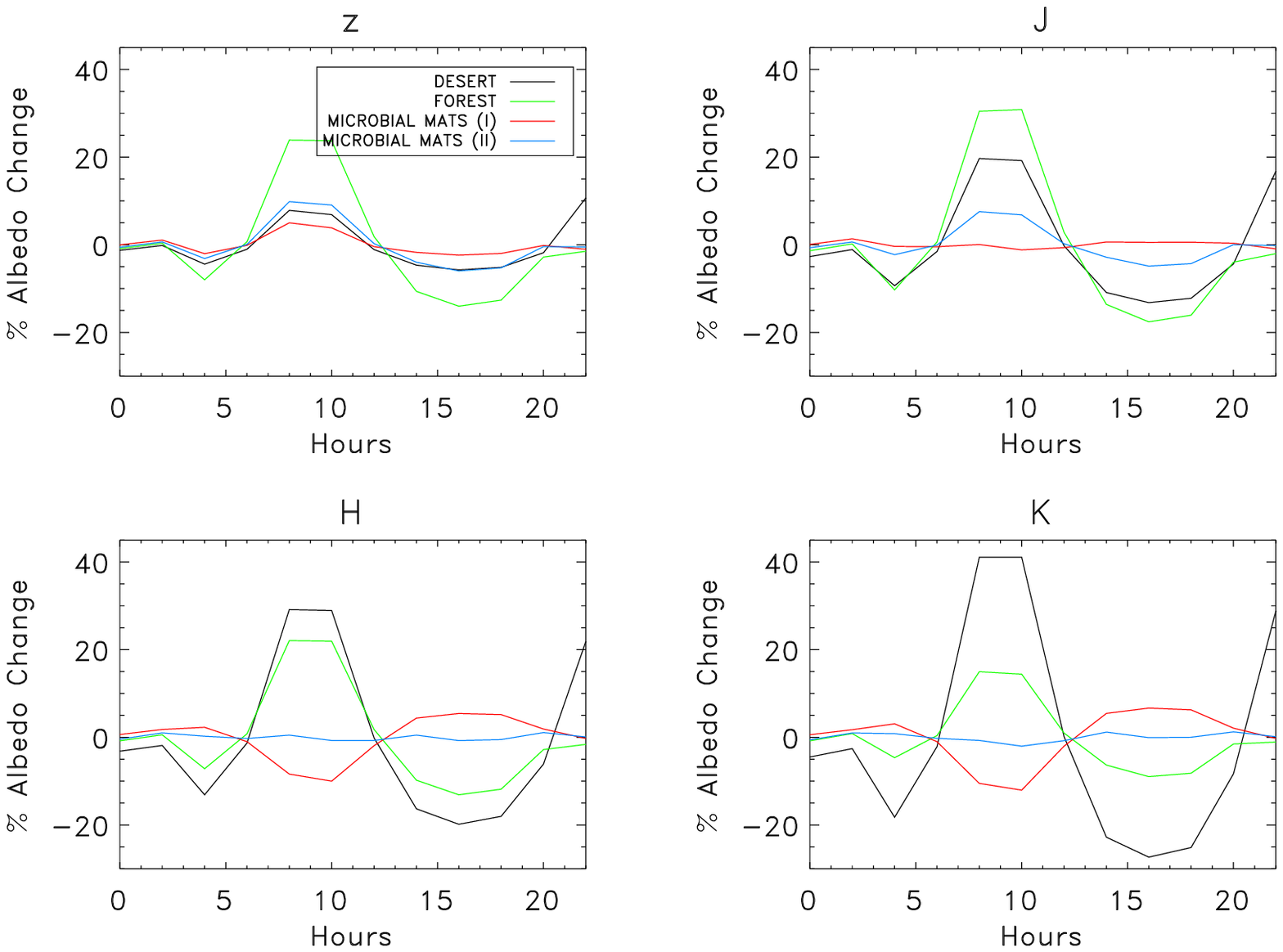}

   \caption{Diurnal light curves of the Earth for the four scenarios contemplated in this paper. The y-scale is the same for all standard photometric filters. The light curves represent the percentage of the albedo change along a day for the 4 scenarios.}
              \label{UBVRI}%
    \end{center}
    \end{figure*}

Figure \ref{UBVRI} shows the photometric time-dependent variations in the disk-integrated reflected light along 
one day. The rotational light curves are plotted for our four scenarios, and for each photometric filter. 
For the Johnson visible system filters, these light-curves show a relatively  smooth shape, with the four 
scenarios being quite similar one to another, and with albedo changes lower than 6\% in general. The only noticeable 
signature is the enhancement in brightness, about 20\%, in the I filter, for the vegetated case (green line) 
between 6 and 12 UT. This is again related to the vegetation's 
signature, the red-edge, since the main continental mass is in full view (and illuminated by the Sun) at this time. 

More interesting results are obtained when one moves redward (bottom part of Figure \ref{UBVRI}). The diurnal 
light curves of a planet with bare desert surfaces (black line) in  J, H and K filters show a considerable 
increase in brightness between 6 and 12 UT of about 20\%, 30\% and 40\% respectively, making immediately obvious the 
existence of a continent. The same is found
in the vegetation case (green line), being these variability of around 30\%, 20\% and 15\% for J, H and K, respectively.

The light curve for a planet with continents covered with microbial mats with cloudiness distribution corresponding 
to vegetation (blue line) is very mutted, with little variability in either the visible or the near-IR spectral 
ranges. In contrast, light curves of the microbial mats case with cloudiness distribution corresponding to desert 
(red line) show a decrease in the albedo of around 10\% in the H, and K filters, when the main continental mass is 
facing the observer. In the J filter these light curves do not exhibit any variation, making it impossible to even discern 
the presence of continental masses. 

\begin{figure*}
   \centering
   \includegraphics{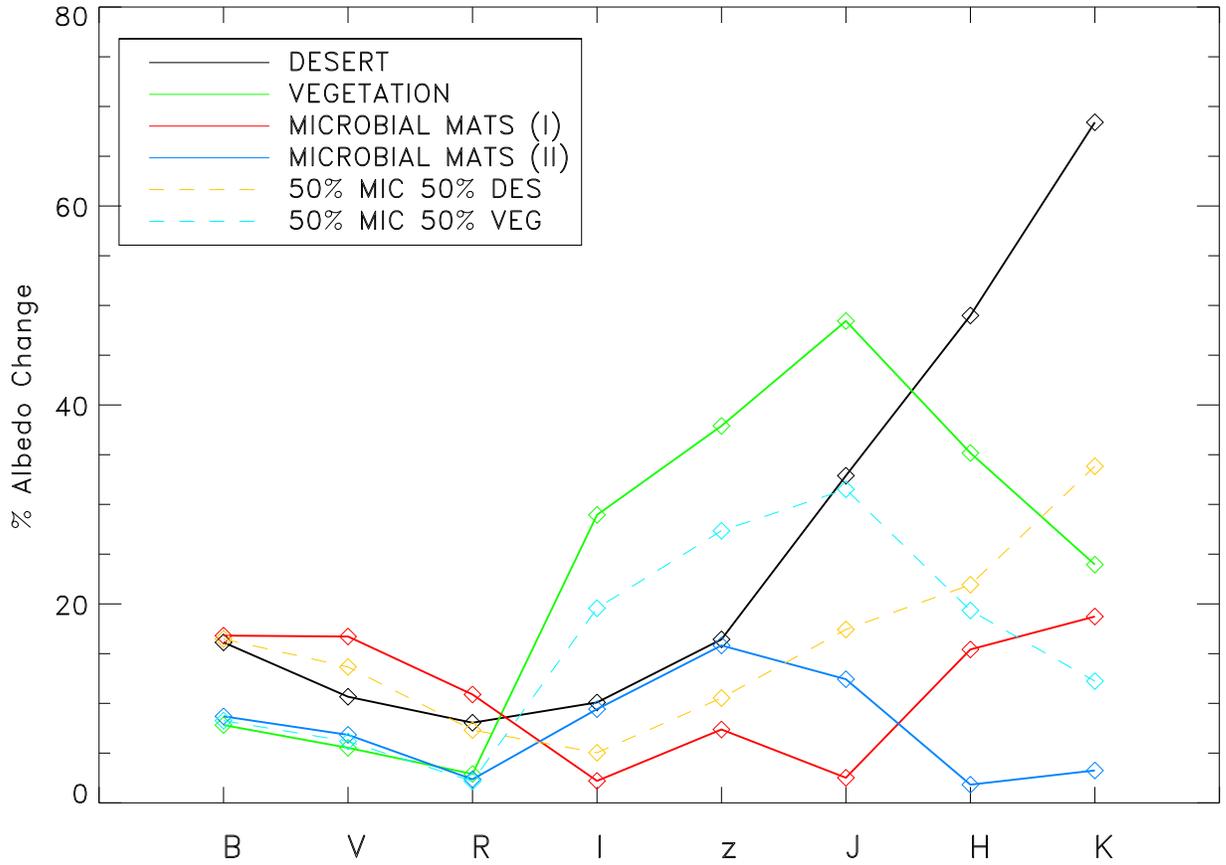}

   \caption{Amplitude of the  albedo variability as a function of the standard photometric filters. Colors
	    represent the different cases studied here: the Earth 500 Ma ago with continents completely covered by deserts 
	    (black solid line), by vegetation (green solid line), by microbial mats (red and blue solid lines) and covered 
	    by a mixture of surface types: 50\% microbial mats - 50\% deserts (yellow dashed line), and 50\% microbial mats
	    - 50\% vegetation (blue dashed line).}
              \label{BVRI_zJHK}%
    \end{figure*}

Figure \ref{BVRI_zJHK} shows the amplitude of albedo variability of each of the studied cases as a function of the different photometric filters. As considering continents completely covered by just only one surface type might not be very realistic, we have also analyzed cases where continents were covered by patchy microbial mats, desert and vegetation. As an example, in Figure \ref{BVRI_zJHK} we present the results obtained for: i) continental masses irregularly covered by $50\%$ deserts and $50\%$ microbial mats, and ii) for continents unevenly covered by $50\%$ vegetation and $50\%$ microbial mats.

The curves in Figure \ref{BVRI_zJHK} show a distinctive shape depending on surface type. When continents are completely 
covered by deserts  (black line) the amplitude of albedo change increases rapidly when one moves from z filter redward, 
while the variability is nearly constant in the visible. When continents are totally covered by vegetation (green line), 
that albedo variability increases nearly monotonously from R filter to J filter, where it peaks, and then decreases redward. 
In contrast to these two cases, when continents are covered by microbial mats (blue and red lines), the amplitude of albedo 
change is not strongly dependent on the filter selected and is much more mutted. Moreover, when we considered that the Earth 500 Ma ago had continents covered by a mixture of microbial mats and deserts/vegetation (yellow and light-blue lines), the shape of these amplitude variability curves resembles the typical shape of the desert/vegetated cases, but with muted variability.

At the light of these results we conclude that it should be a priori possible to determine the type of continental 
surface on a rocky Earth-like planet based on color photometry and given a high enough signal-to-noise ratio. A planet 
covered by bare desert areas will have a large diurnal variability  in the J, H and K bands. While this is also the case for vegetated continents, in this case an additional bump 
in the I band is to be expected, which would be missing in a desert planet. Furthermore, the hypothesis of 
deserted continents could be further tested/confirmed by observations in the mid-IR photometric bands (\citealt{Gom12}). 

In the case of a planet with extended microbial mats, its nature would be determined by the 
detection of the presence of continents by using the light curves in B and V bands, and then finding a lower than expected 
percentage variability in the J, H and K bands, with the reflection peak characteristics of plants and deserts missing.  

Although not shown in this work, we also studied the same scenarios but with cloud-free atmospheres. As expected, we find that the omission of clouds decreases the amount of reflected light in comparison to the respective non-clear sky condition. Nevertheless, daily variations in the disk-integrated spectra become more dramatic in the cloud-free case than 
in the cloudy one. These cases, however, are highly unrealistic.

%__________________________________________________________________

\section{CONCLUSIONS}

In this paper we have used a radiative transfer code to simulate the globally-integrated spectral variability of 
the Earth 500 Ma ago, using four possible scenarios regarding the continental surface properties. Our simulations also include 
realistic distributions of clouds over the whole planet at three altitude layers. We find that 
as the continental surface  changes from desert ground to microbial mats and to land plants, it produces detectable 
changes in the globally-averaged Earth's reflectance that vary substantially as the Earth rotates. By binning the data into standard astronomical photometric
band we see that the variability of each surface type is located in different bands and can induce reflectance 
changes of up to 40\% in periods of hours. We conclude that 
using photometric observations of an Earth-like planet at different photometric bands, it would be possible to 
discriminate between bare continental surfaces, large microbial mats extensions, or plant-covered continents. 
While in the recent literature the red edge feature of vegetation at visible wavelengths has been proposed as a signature for land plants, observations in near-IR bands can be equally suited for this purpose.

%__________________________________________________________________

\acknowledgments

The authors thankfully acknowledge the technical expertise and assistance provided by the Spanish
Supercomputing Network (Red Espa\~nola de Supercomputaci\'on), as well as the computer resources used:
the La Palma Supercomputer, located at the Instituto de Astrof\'isica de Canarias. 
The authors also acknowledge support from the Spanish MICIIN, grant CGL2009-10641 and AYA2010-18080.

%======================================================================================================

%------------------------BIBLIOGRAFIA------------------------------------------------------------------

%======================================================================================================

% \bibliographystyle{aa}
% \bibliography{mesr_biblio}

%======================================================================================================
%------------------------FIGURAS MIAS-----------------------------------------------------------------------
%======================================================================================================

% % %    \begin{figure*}
% % %    \centering
% % %    \includegraphics{500Ma_red_edge.eps}
% % %    \caption{Same as in Figure \ref{500maespectro}, however only the spectral region near the vegetation's red-edge 
% % % 	    (0.60-0.95 $\mu$$m$) is plotted, and figure panels are grouped by
% % % 	    time of the day instead of scene type. Cases where continents are covered by deserts are 
% % % 	    plotted in black, by forest in green, by microbial mats with cloudiness
% % % 	    information corresponding to desert in red, and microbial mats with cloudiness information 
% % % 	    corresponding to vegetation in blue.}  
% % %                                                                                          
% % %               \label{500marededge}%
% % %     \end{figure*}

\begin{table}
\begin{center}
\caption{The diurnal variability of the vegetation's red edge strenght for the Earth 500 Ma ago.
\label{tbl-1}}
\begin{tabular}{crrrrr}
\tableline\tableline
Surface Type & 2 UT & 8 UT & 12 UT & 18 UT & \\
\tableline
Desert  						&1.14 &1.35 &1.16 &1.03\\
Forest   	 					&1.03 &1.05 &1.03 &1.01\\
Microbial Mats (I)			 		&1.05 &1.12 &1.05 &1.02\\
Microbial Mats (II) 					&1.04 &1.10 &1.05 &1.02\\
\tableline
\end{tabular}
%% Any table notes must follow the \end{tabular} command.

\tablecomments{Ratio between the intensity in the 0.740-0.750 $\mu$$m$ \\ and in the 0.678-0.682 $\mu$$m$ range.}
\end{center}
\end{table}

% % \begin{table}
% % \begin{center}
% % \caption{Percentage of the Daily Variability and Mean Albedo of the Light Curves for Each Epoch.
% % \label{tbl-1}}
% % \begin{tabular}{crrrrr}
% % \tableline\tableline
% % Surface Type & 2 UT & 8 UT & 12 UT & 18 UT & \\
% % \tableline
% % Desert  						&0.88 &0.74 &0.86 &0.97\\
% % Forest   	 					&0.97 &0.95 &0.97 &0.99\\
% % Microbial Mats (I)			 		&0.96 &0.91 &0.95 &0.98\\
% % Microbial Mats (II) 					&0.95 &0.89 &0.95 &0.98\\
% % 
% % \tableline
% % \end{tabular}
% % %% Any table notes must follow the \end{tabular} command.
% % 
% % \tablecomments{These quantities have been calculated by performing the mean of these results obtained
% % for 2000 January, March and July.}
% % \end{center}
% % \end{table}

\end{document}